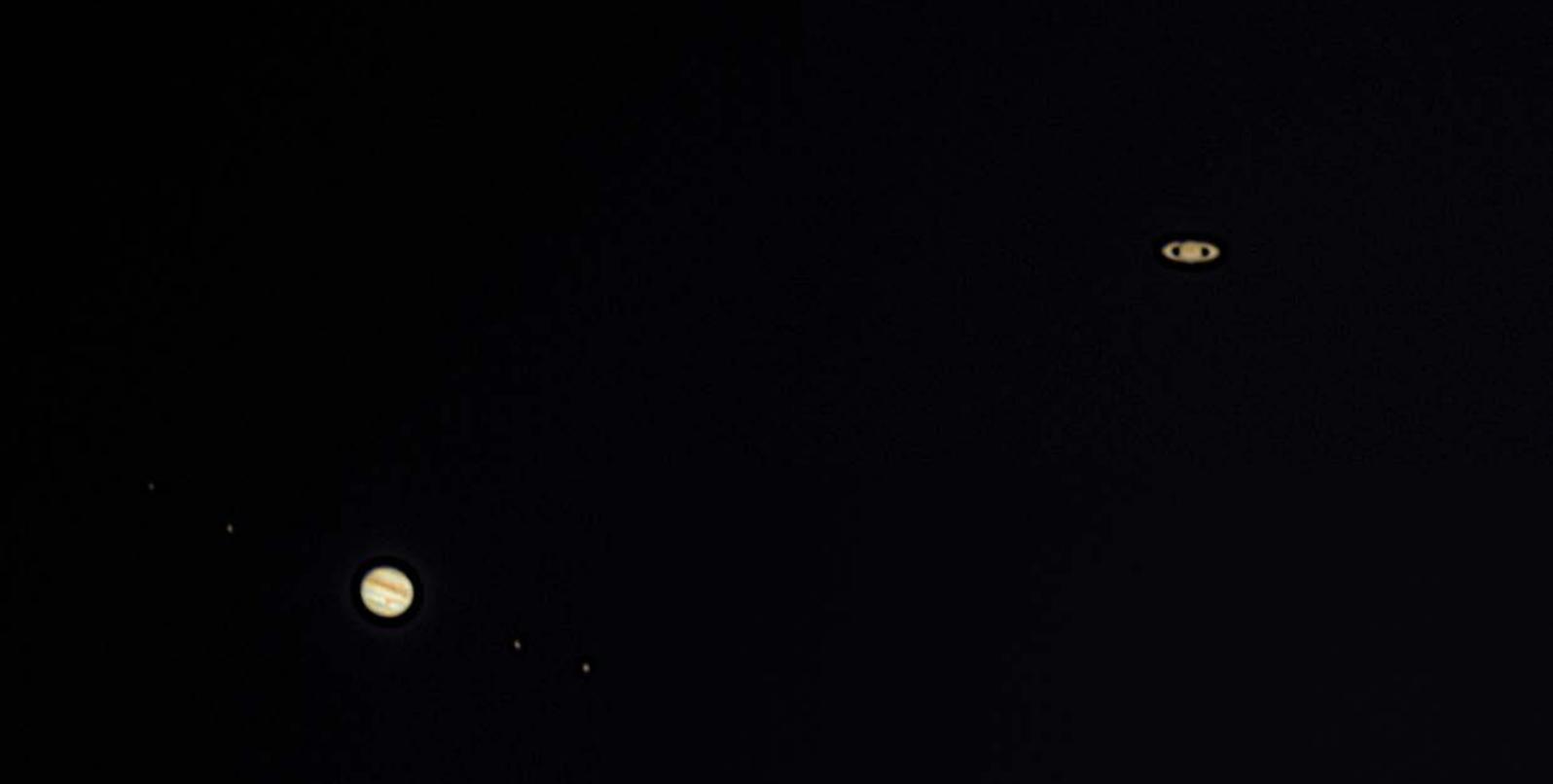

# Sharing the great conjunction

**Claire Davies** recounts a successful outreach project based on Saturn and Jupiter's 2020 conjunction, a celebration of closeness in a year dominated by distancing.

The great conjunction of 21 December 2020 saw Jupiter and Saturn appear together in the sky, separated by just a tenth of a degree (equivalent to a distance five times smaller than the diameter of the full Moon). This provided a potential once-in-a-lifetime opportunity to view the solar system's two biggest planets – and up to five of their moons – through a telescope eyepiece at the same time (figure 1). Moreover, this was the first such opportunity, ever; previous observable conjunctions at similarly close separations took place before the development of the telescope in the early 1600s.

Our team of scientists from the University of Exeter's Astrophysics Group and Exeter Science Centre worked with local social enterprises to develop a series of promotional and concurrent events to tie in with our live telescope broadcast of Jupiter and Saturn, to celebrate this spectacular celestial event. We hoped not only to inform and educate the public about great conjunctions, and the solar system more generally, but also to bring some light relief in what had been a rather difficult year.

## Opportunity for outreach

The public engagement opportunity provided by the great conjunction was initially identified by Matthew Bate of the University of Exeter, who pitched the idea to the Astrophysics Group and Exeter Science Centre in late summer/early autumn 2020. The Stellarium app showed that Jupiter and Saturn would be visible from Exeter for up to 90 minutes after sunset. We realized that the early December sunsets, and the timing of the conjunction within the school winter break, would make the event particularly useful for engaging young children with astronomy.

**1** *Jupiter and Saturn close to conjunction on 22 December 2020. Jupiter's four Galilean moons are also visible.*
(Image by Steven Rieder using a ZWO ASI224MC camera on a Sky-Watcher 200 mm f/5 Newtonian telescope. Composite of 1000 frames, stacked and processed with AutoStakkert, Registax and Adobe Lightroom)

*"We had a week-long window of opportunity in which our live stream could take place"*

Early suggestions for activities included in-person public observing events with telescopes set up on the University of Exeter's Streatham campus or, alternatively, spread across locations around Exeter and neighbouring Exmouth. However, we were keen for as many people to witness the event with us as possible and, based on attendance figures at similar past events, the (optimistic) 90-minute viewing window would not have provided much time for each attendee to view the conjunction. Moreover, as time went on and a second lockdown to slow the coronavirus pandemic was looking increasingly likely, we chose to abandon the idea of a physical event. Instead, armed with newly gained confidence in making video presentations thanks to moving our lectures and academic conferences online, we focused on hosting a live telescope feed. This would be broadcast via the existing "Physics at Exeter" YouTube channel.

Due to the likelihood of poor weather in the UK in December, we were disinclined to organize an event on a single fixed date. Fortunately, Stellarium had shown that Jupiter and Saturn would remain close enough to appear within the same telescope eyepiece for a few days either side of 21 December. This provided a helpful week-long window of opportunity in which our live stream could take place. To advertise this effectively, we realized that we would need both a way of securely collating contact details for interested potential attendees and a reliable alert mechanism to announce our chosen date at relatively short notice.

## Going online

Sam Morrell designed and built a website – jupitersaturn2020.org – which we used to collate information explaining what the 2020 great conjunction was as well as how, where and when to view it. In particular, we promoted our live broadcast and embedded a link to a Google Form which we used to collate a GDPR-compliant mailing list. We also chose to prepare educational digital

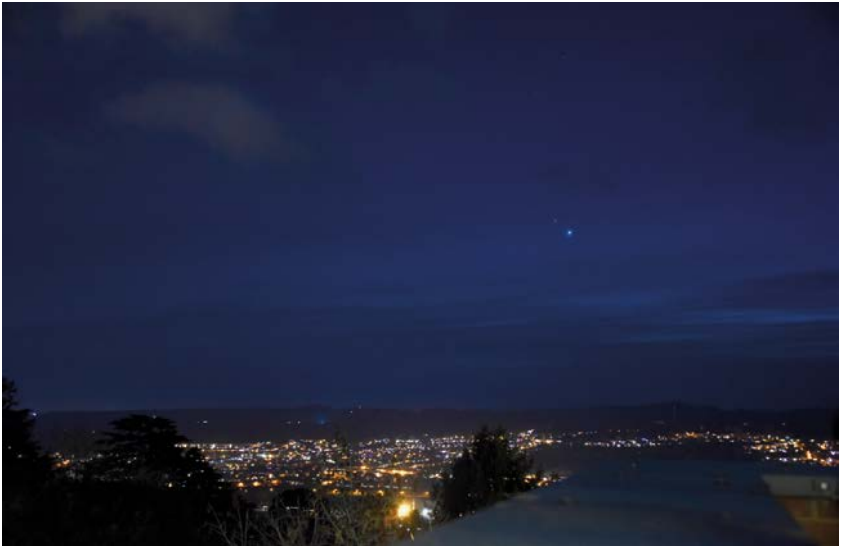

2 *The slab block roof of the Physics building, with its unobstructed views of the south west sky, provided an ideal location from which to broadcast the live telescope feed.* (Matthew Bate) content to encourage traffic to our website and increase the reach of our event promotion. This comprised a series of videos designed to educate viewers about great conjunctions, the solar system and the motions of the planets, as well as promoting research conducted by members of our Astrophysics Group on planetary atmospheres and the formation of planetary systems. These were hosted on the "Physics at Exeter" and "Exeter Science Centre" YouTube channels and embedded on our website.

Our first video, prepared and delivered by Bate, explained when and where to look to view the great conjunction as well as what people could expect to see with a small telescope, a pair of binoculars or the naked eye. The timing of this video ended up being fortuitous: a few weeks after it was uploaded on 16 November, an up-tick in searches for content associated with the great conjunction resulted in this video being featured on YouTube users' personalized home pages, suggested videos and other browsing features. The video went viral: by early December it was gaining views at a rate of around 10 000 per day and, as of January 2021, has acquired over 632 000 views (and almost 10 000 likes). Around three-quarters of the total traffic to this video resulted from promotion by YouTube.

Our team prepared and uploaded additional content throughout late November and early December. A video aimed at young children, written and produced by PhD student Federica Rescigno, provided context on the novelty of this celestial event while presenting interesting facts about Jupiter and Saturn. Viewers were also encouraged to join in with Federica as she cut out and coloured in different sized circles of paper to demonstrate the difference in size between the Earth and Jupiter. Exeter Science Centre's directors, Alice Mills and Natalie Whitehead, recorded interviews with Nathan Mayne and Stephen Thomson of Exeter University about their research. Our digital content was rounded off with videos on the history of great conjunctions (written and produced by Bate) and on the "grand tack" theory of planetary migration in the early solar system (produced by Sebastiaan Krijt with assistance from Elizabeth Tasker [Japanese Aerospace Exploration Agency, JAXA] and Sean Raymond [Laboratoire d'Astrophysique de Bordeaux]).

Our live broadcast was designed to follow a *BBC Stargazing Live* style multi-stream whereby hosts would introduce the show and hand over to different guest experts. We opted for this format as it would enable us to run with the telescope view while the weather was clear and switch to informative pieces if and when cloud rolled in, or the presenters needed a break. Additional content based on questions submitted by schoolchildren to the Exeter Science Centre website was prepared ahead of the live broadcast. This included a short video produced by Jenny Hatchell, which showed the relative locations of the Earth, the Sun, Jupiter and Saturn during the conjunction. Steven Rieder and Chris Brunt, together with Hatchell and Bate, prepared the telescope set-up and acquired back-up footage of Jupiter and Saturn during clear evenings in the run-up to the live broadcast, all the while abiding by the social distancing regulations in place at the time. Whitehead and Mills led the development of a script to be used on the night, while Morrell worked tirelessly to ensure the broadcast could transition smoothly between different live and prerecorded feeds. This preparation allowed us to practise delivery ahead of the live stream, including ironing out handovers, which had to overcome a delay of a few seconds between delivery and broadcast (figure 2).

### Community engagement

I have enjoyed previous success organizing public observing and stargazing events with Exeter-based social enterprises. These employ a collaborative engagement model, where core learning themes are discussed alongside a community's needs and assets to develop events. Such existing links with local communities were recognized as beneficial to our event, providing us with methods of fostering local support and building up local interest. Our event was also seen reciprocally as an asset to local communities, with our potential to provide a positive impact on community well-being through social connectivity and the relative ease with which bright celestial objects appearing low in the sky could be tied in with the magic of Christmas.

Our major partner was Interwoven Productions CIC, which encourages citizens to commission the learning they want, on their terms, through its unique Squilometre technique (the five current Squilometre neighbourhoods in Devon represent around 20 000 citizens). Interwoven works hard to maintain constant, two-way communication with its communities (and even managed to keep this up during the coronavirus pandemic) and, as such, it represents a rich and receptive conduit to community engagement.

Interwoven also supports a network of participatory arts practitioners. We worked with Interwoven associate artist (and specialist in immersive, educational theatre) Boo to a Goose Theatre to redevelop their *Space Cats* production (figure 3). This Arts Council-funded sensory play, aimed at primary-school children aged eight and under, had originally been developed to commemorate the 50th anniversary of the Apollo 11 Moon landing and had already visited 30 venues including schools, nurseries and libraries. The play features a cat-loving grandad, Houston

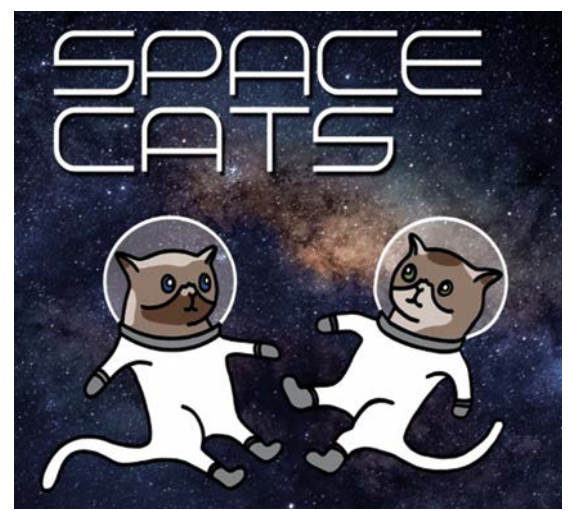

3 *Boo to a Goose Theatre's* Space Cats *show was used to educate and inform young audiences about the solar system while the accompanying activity sheets explained to teachers, parents and carers how to view the great conjunction.* (Julian Hoad, Another Planet design)

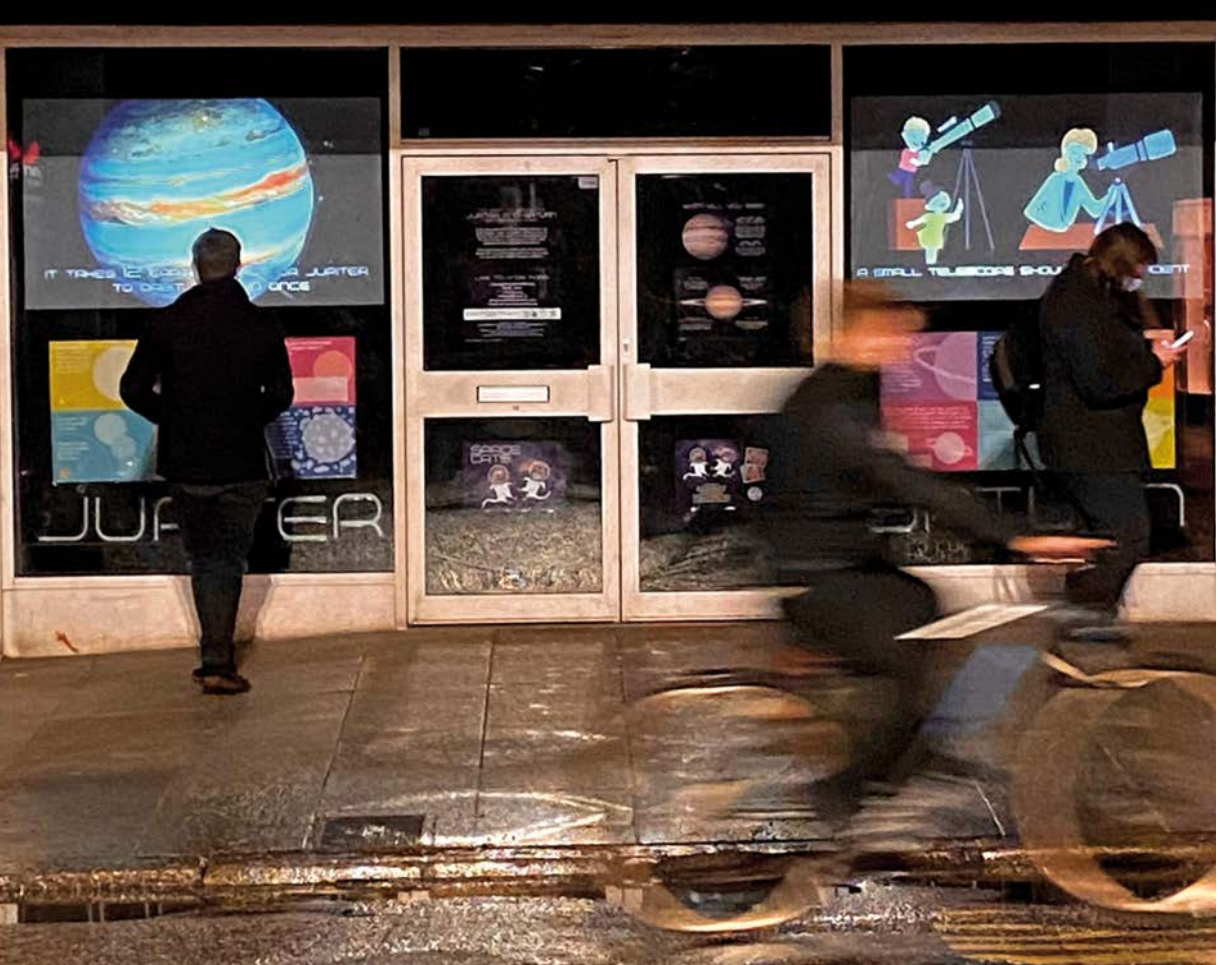

**4** *The city-centre shopfront display on Sidwell Street, Exeter, helped to promote the community engagement activities and advertise the live broadcast of the great conjunction.* (Julian Hoad)

Collins, and his two curious kittens who discover a spaceship belonging to female rocket scientist Prof. Agena B Armstrong and fly off to the Moon. Mills and I consulted with Boo to a Goose Theatre to include information about Jupiter and Saturn and the great conjunction in *Space Cats*; this adapted version of the show visited local schools during December, reaching 180 children across Key Stage 1 groups in schools in Exeter and Exmouth. Interwoven's Julian Hoad and Anna Matthews designed accompanying activity sheets, which were distributed at the shows to provide parents and carers with details of how to view the great conjunction, including how to access our live broadcast.

Julian and JoJo Spinks of Interwoven were also keen to use this occasion to bring astronomy and science into the heart of Exeter in an accessible and Covid-safe manner (figure 4). This idea developed into a week-long dynamic city-centre shopfront installation, which was active from 16–23 December. Bespoke videos and infographic posters, based on the content we produced for our website, were created by Interwoven associate artist Dom Lee. The two videos were kept short to avoid viewers having to stand out in the cold for long periods, while the colourful posters were designed to be appealing to children and placed low in the shop windows to catch their attention. Posters on the doors of the shop advertised our community engagement activities and how to view our live broadcast. We were pleased to see the shop's location enjoy significant footfall throughout the period of the installation, with an average of 65 people passing by every five minutes. This was no doubt helped by Exeter remaining in tier 2 local restrictions throughout December. Our monitoring indicates that our shopfront achieved at least 3000 separate engagements with individuals, families, couples and groups of young adults. We were also pleasantly surprised to see that, with Sidwell Street acting as a temporary bus depot while Exeter's bus station is redeveloped, the shopfront enjoyed a captive audience of bus drivers and passengers awaiting driver handovers.

Concurrent observing events were organized by Interwoven and by PRISM Exeter (an LGBTQ+ STEM network). The former was organized as an in-person event targeted at residents of Exeter's Burnthouse Lane council estate, who were encouraged to book scheduled slots to have a go at viewing the "Christmas star" through binoculars on 20 December. For most attendees, this was a novel opportunity and prompted further discussion about the solar system and astronomy. PRISM Exeter held virtual social events on 20 and 21 December targeted at Exeter's LGBTQ+ community. Attendees were guided through observing Jupiter and Saturn and they quizzed local LGBTQ+ astronomers about what was happening and why. Our live broadcast was also shared with attendees at both events.

## The main event

We announced our chosen date – 20 December 2020 – in an email delivered to more than 8400 people, who had signed up to our mailing list. 5600 of these opted in to receive notice of any future events that we hold. Our live broadcast attained a peak concurrent viewing figure of 3666. This was helped, in no small part, by the roughly 200 mentions our event received across international media, including TV and radio interviews with Sky, the BBC and Al Jazeera; and it being listed as one of six webcasts of the great conjunction by space.com (space.com/great-conjunction-winter-solstice-2020-jupiter-saturn-webcasts). During our live broadcast, PhD student Adrien Houge monitored the YouTube chat and selected questions to pass to our presenters. This allowed answers to be prepared behind the scenes before they were delivered live. Messages in the YouTube chat peaked at 63 per minute during the broadcast. The live broadcast has since been viewed more than 300000 times, receiving 3000 likes. Together, the YouTube videos and live-stream event increased the number of subscribers to the "Physics at Exeter" YouTube channel from around 150 in mid-November to more than 8500. The problem facing us now is how to maintain this audience but, as we are all quick to admit, that is a rather nice problem to have. ●

**AUTHOR**

Dr Claire Davies is a research fellow in the Astrophysics Group, University of Exeter, and founder of PRISM Exeter

**ACKNOWLEDGEMENTS**
The Great Conjunction live broadcast and associated activities were devised by: members of the **University of Exeter Astrophysics Group** – Matthew Bate, Chris Brunt, Claire Davies, Jenny Hatchell, Adrien Houge, Sabastiaan Krijt, Nathan Mayne, Sam Morrell, Steven Rieder, Federica Rescigno and Stephen Thomson; **Exeter Science Centre** directors – Alice Mills and Natalie Whitehead; and **Interwoven Productions CIC** – Marie Cassidy, Mark Cassidy, Julian Hoad, Dom Lee, Anna Matthews and JoJo Spinks. The *Space Cats* and shopfront installation elements were supported by funds awarded by the Institute of Physics South West branch and Resolute Photonics Ltd to Interwoven Productions CIC. PRISM Exeter's activities throughout 2020 were supported by a Diversity in Science Grant awarded by the Biochemical Society to Dr Claire Davies